# Observation of new quantum interference effect in solids

Avto Tavkhelidze *, Amiran Bibilashvili and Larissa Jangidze
*Tbilisi State University, Chavchavadze Ave. 13, 0179 Tbilisi, Georgia*

Alex Shimkunas and Philip Mauger
*Nanostructures, Inc., 3070 Lawrence Expressway, Santa Clara, CA 95051*

Gertrude F. Rempfer, Luis Almaraz, Todd Dixon
*Portland State University, 220 Science Building 1, 1825 SW Broadway, PO Box 751, Portland, OR 97207*

Martin E. Kordesch
*Clippinger 158, Department of Physics, Ohio University, Athens, OH 45701-2979*

Nechama Katan and Hans Walitzki
*Avto Metals plc, London England*

In order to achieve quantum interference of free electrons inside a solid, we have modified the geometry of the solid so that de Broglie waves interfere destructively inside the solid. Quantum interference of de Broglie waves leads to a reduction in the density of possible quantum states of electrons inside the solid and increases the Fermi energy level. This effect was studied theoretically within the limit of the quantum theory of free electrons inside the metal. It has been shown that if a metal surface is modified with patterned indents, the Fermi energy level will increase and consequently the electron work function will decrease. This effect was studied experimentally in both Au and $SiO_2$ thin films of special geometry and structure. Work function reductions of 0.5 eV in Au films and 0.2 eV in $SiO_2$ films were observed. Comparative measurements of work function were made using the Kelvin Probe method based on compensation of internal contact potential difference. Electron emission from the same thin films was studied by two independent research groups using Photoelectron Emission Microscopy (PEEM).

* E-mail address: avtotav@geo.net.ge

# Introduction

The wave properties of electrons inside a solid are well known and understood. There are some nanoelectronic devices, such as resonant tunneling diodes and transistors, super lattices, quantum wells and others that are based on the wave properties of the electron [1]. Under certain conditions an electron in a solid can be regarded as a planar wave. The main requirement that should be satisfied is that at least one dimension of the solid should be equal to or less than the mean free path of the electron inside the solid. In this case, the electron can move without scattering and could be regarded as de Broglie wave. It is difficult to satisfy this requirement because the electron mean free path in most solids is in the range of 1-10 nm at room temperature. Transport properties of solids (current and heat transport) are defined by electrons, having energies close to the Fermi level, and the mean free path is given for those electrons. Other free electrons inside solids, for example electrons having energies below Fermi level in metals, do not participate in current and heat transport, because it is quantum mechanically forbidden for them to exchange energy with the environment (all quantum energy levels nearby are occupied), and hence the mean free path of such electrons is formally infinite. Such electrons will remain ballistic inside relatively large structures [2]. In this work we use wave properties of such electrons to change electronic structure of solid in the way, that work function of solid could be reduced and regulated precisely. Such materials will find many applications in devices based on electron emission, electron tunneling [3] and in semiconductor industry.

Assume a solid with the surface geometry, as shown in Fig 1a, in which periodic indents are introduced in the flat surface of the solid. Let us consider an electron traveling towards the border of the solid as planar wave 1. Wave 1 will reflect back from the border of the solid because the electron does not have enough energy to leave the solid. Because of the geometry of the surface there will be two reflected waves. One will reflect from the top of the indent (wave 3) and other will reflect from the bottom of the indent (wave 2). If the indent depth is one quarter of the de Broglie wavelength of the electron, waves 2 and 3 will interfere destructively and there will be no reflected wave. As a result, an electron of certain energy cannot reflect back from the surface because of its wave nature. On the other hand, the electron cannot leave the solid and enter the vacuum because it does not have enough energy to overcome the potential barrier. For obvious reasons, the electron cannot simply stop near the surface either [3]. From the quantum mechanical point of view, we can say that all possible final quantum states for that particular electron are forbidden. As all the final quantum states are forbidden, then the initial quantum state is also forbidden. As a result, the density of the quantum states inside the solid will be reduced. A 3D drawing of the solid is shown on Fig. 1b. If we regard the solid as potential energy box, there will be standing de Broglie waves inside the solid. Each standing wave corresponds to the quantum state which could be occupied by the free electron. The number of standing waves inside such a 3D structure is lower than in the case in which there were no indents and all the walls of the solid were plain.

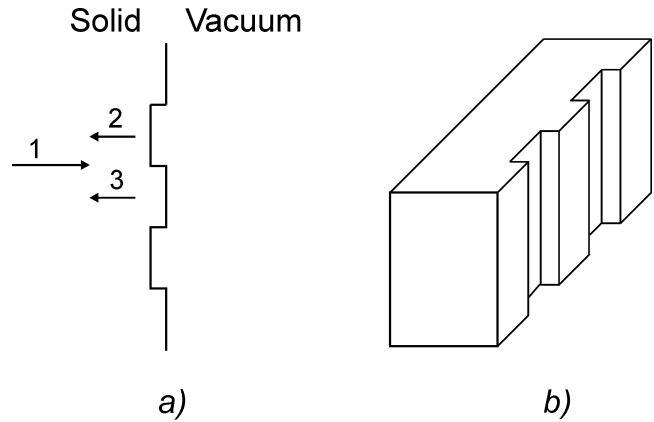

Fig. 1. a) de Broglie wave interference diagram. b) Geometry of modified potential energy box.

A theoretical analysis that starts with the Schrödinger equation and then calculates the density of the quantum states and diameter of the Fermi sphere in k space shows that the density of the quantum states is dramatically reduced when indents are introduced [4, 5]. The calculation was made within the limit of the theory of free electron gas in the metal (later a similar method was used in [6]), and the result is shown on Fig. 2. The indents in the wall cause the density of quantum states to be reduced in whole energy region below the Fermi level (Fig.2b). Once the number of quantum states is reduced, given that there is no reduction in the number of free electrons, electrons are forced to occupy higher energy levels. As a result we have an increase of the Fermi level and a corresponding reduction of the work function.

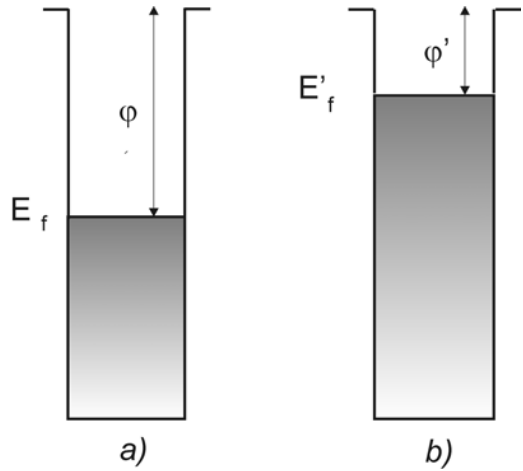

Fig. 2. Energy diagram of a) metal having plain walls, b) metal having periodic indents on one of the walls.

Some authors have named the effect as the Avto effect (note made by Avto Tavkhelidze).

In practice there are two limiting factors in achieving the Avto effect. First, the surface roughness should be less than the de Broglie wavelength of the electron, in order to avoid the scattering of de Broglie waves on the surface and reducing the effect. Secondly, it is ideal for the solid to be amorphous or single crystalline in order to allow electrons to remain ballistic while moving between the indented wall and the opposite wall. Polycrystalline structure of the material will destroy the effect because of scattering of the electrons on the grain boundaries.

It was the objective of this work to reduce the work function and correspondingly increase the emission from various solids, in the way of modifying geometry of the surface of the solid. Materials with reduced work function will find broad applications in devices working on the basis of electron emission and electron tunneling. Besides it, such a materials will be useful in semiconductor industry, particularly for the structures in which contact potential difference between two layers play important role.

## Sample preparation and experimental results

At Tbilisi State University (TSU) thin gold films, having indents on both sides, were prepared to observe the Effect (Fig.3a). Gold was the material of choice because it does not form natural oxide on the surface and allows exposure of the films to the atmosphere. Au film was deposited on a $Si/SiO_2$ (dry thermal oxide) substrate, and, after a conventional cleaning procedure, a layer of photoresist S1813 of thickness 0.4 μm was attached at 4000 rpm (photoresist was solved prior to attachment). Optical microscope MII-4 was used for thickness control. Periodic lines 0.8 μm wide were created in the photoresist using UV photolithography, and the $SiO_2$ was etched using $NH_4F + HF + H_2O$, at a rate of 1 nm/sec to a depth of 10-50 nm. In the next step, the photoresist was removed using acetone followed by a conventional cleaning procedure. A further layer of photo resist was attached, and another photolithographic step was used to form large structures using a lift-off process. The substrate was then placed in deposition chamber, and after a vacuum $10^{-6}$ Tor was obtained, heated to 80 C to remove water from the surface. The substrate was cooled to temperatures between -16 C and -22 C, and a thin film of 2-3 nm thickness, evaporated from a mixture of Au and Cr, was deposited on $SiO_2$ to form an adhesion layer (not shown on fig3a for simplicity). Following deposition, the wafer was moved rapidly (maximum 5 s) to another location, where Au film of thickness of 60 nm was deposited using fast thermal evaporation of Au wire (99.999% purity). The substrate was heated up to room temperature and the wafer was taken out of deposition chamber. The final step was a conventional lift off process to form large structures.

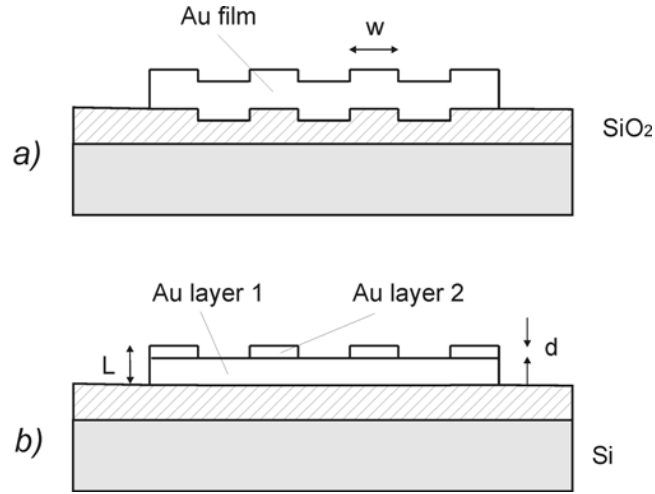

**Fig. 3.** Schematic cross-section of test samples made at TSU (a) and Nanostructures Inc. (b). Adhesion layers between $SiO_2$ and Au film are omitted for simplicity.

Measurements of the work function were made using Kelvin Probe (KP) method. All measurements were comparative to exclude absolute inaccuracies: the difference between KP reading on a flat region of the gold film was compared to the reading from the indented region of the film. The structure of the films was analyzed using X-ray diffraction fig.4.

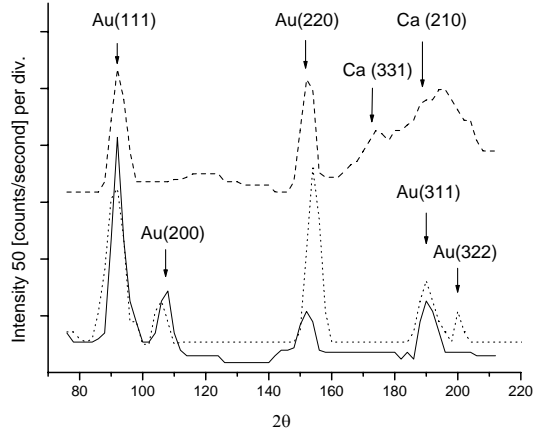

Fig.4. X-ray diffraction patterns (Co source 1.78 A): Solid line-Au film deposited on cold substrate; Dot line –Au film deposited on room temperature substrate; Dash line – Au film containing Ca atoms.

For all samples measured, the indented regions showed a reduced work function (WF) compared to the flat regions. The magnitude of this reduction of WF depended on the structure of the gold film and the depth of the indents. Films having an amorphous structure show much higher reduction in WF than films having a polycrystalline structure. All polycrystalline films show a WF reduction less than 0.1 eV whilst for amorphous films the reduction of WF is in the range of 0.2-0.5 eV. Amorphous films were obtained using deposition of gold on cooled substrate (as described above), and polycrystalline ones were obtained by deposition of gold on room temperature substrate fig.4. No other technological parameter except substrate temperature was different in the two deposition processes. The difference in WF was more pronounced for samples that were deposited in a cleaner environment (by plasma cleaning the deposition chamber prior to deposition), because for the cooled wafer the residual gas pressure and composition has considerable influence on the structure of the film. The difference in WF reduction up to 10 times, shows that structure of the film has principal importance for observation of Avto effect. This experimental result is in full agreement with theory.

One further unplanned experiment confirmed the importance of the structure of the film. When the Au film was fabricated in a deposition chamber previously used for Ca, it unexpectedly showed a reduction of WF of 0.06 eV instead of the expected 0.5 eV. Subsequent X-ray analysis fig.4 revealed the presence of Ca atoms inside the Au film, and also the Au film was polycrystalline instead of amorphous. It was obvious that Ca contamination changed the film structure to polycrystalline, resulting in the Effect almost vanishing. After the deposition chamber was cleaned, new samples were fabricated which showed better amorphousness on X-ray analysis, and the WF difference increased to 0.2 eV. Fabrication of samples following further cleaning of the chamber, by dismantling followed by chemical and mechanical treatment to remove the thin layer of Ca completely from the parts of deposition chamber, yielded Au films showing a WF reduction of 0.4 eV and having amorphous structure by X-ray analysis.

It was observed that the strength of the effect depends on the depth of the indents. Samples having Au film thickness of 60 nm, and indent depth of 50, 20 and 10 nm, show WF reduction of 0.16 eV, 0.25 eV and 0.56 eV respectively. This experimental result is in quantitative agreement with the prediction of the theory. In theory Fermi level increases as $E_f^d = E_f^0 (L/d)^{2/3}$ where $E_f^d$ is the Fermi energy of the indented film, $E_f^0$ is the Fermi energy of plane film, L is the thickness of the film and d is the depth of the indent (formula could be used for real sample description only for the case d>Ra, where Ra is surface roughness). Formula shows that the less d is (for L=constant), the more increase of Fermi level and consequently the reduction in WF (fig.2) should be, and the same thing is observed experimentally. Other possible explanations of the observed dependence could be the increase of the roughness of the indents in $SiO_2$ films with increasing etching depth, or the increase of under photoresist etching with increasing indent depth. AFM measurements show that there is no increasing roughness with increasing indent depth, but there is some increase of under photoresist etching with increasing etching depth.

Reduction of WF of 0.2 eV was observed on indented area of $SiO_2$. The measurement was made on several samples and yielded almost identical results.

Test samples in Nanostructures Inc. were prepared using conventional optical lithography and etching techniques to demonstrate the work function reduction of indented gold surfaces. The sample geometry was a little different from ones made in TSU. Indents (more precisely protrusions) in Au were made only on the surface of the film, while film base remained flat (fig.3b). First 50 nm Au film was deposited on plane Si/SiO$_2$ substrate at room temperature (layer 1). Next photoresist was attached to the gold film and 0.8 um lines were opened. Afterwards 7-10 nm thick gold film was deposited at room temperature (layer 2). Finally lift off of second layer of gold was made to form periodic lines. AFM was used to verify the height, width, and period of the protrusions, and to evaluate the surface and edge roughness of the indented patterns.

WF reduction of 0.1 eV was observed in these samples. Low value of WF reduction could be explained by the fact that Au films were deposited on room temperature substrate. Another explanation is the difference in geometry. As mentioned above films had indents on one side only, unlike TSU films having indents on both sides.

Measurements made on one of the Nanostructures sample provided more information for understanding the Avto effect. One of the samples was prepared so that it contained five areas each having different surface geometry. The first area was plain and the other four areas had various indent widths of 0.8, 0.6, 0.4, 0.2 microns, but the same depth and length. Comparative measurements made between 1-2, 1-3, 1-4, 1-5 show that WF difference increases with reduction of indent width. Theory does not predict such dependence in the case of ideal structure of the film (i.e. single crystal or amorphous). But in reality the film will always contain crystals, and the electron mean free path will be limited, even for electrons below the Fermi level.

In PSU images were collected using high-resolution photoelectron microscope [7]. Fig 5. shows PEEM image of 16 squares of dimensions of 20x20 micron each. Some squares have indented surfaces and others have plain surfaces. Indented and plain areas are placed like chess desk. Intensity on the image is proportional to the electron emission. Indented areas appear brighter, corresponding to more electron emission and consequently to less WF.

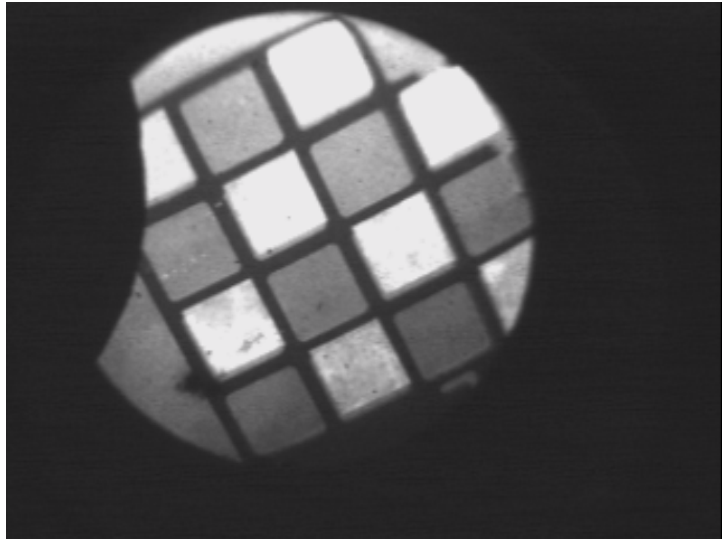

Fig. 5. PEEM image of Avto effect. Rectangular areas placed like chess desk. Bright squares correspond to indented areas exhibiting Avto effect and darker squares corresponding to flat gold film.

In Ohio University the PEEM images were collected using a Bauer-Telieps style LEEM/PEEM instrument [8]. The acceleration voltage was 15 kV, the illumination source an HBO 100 mercury short arc lamp incident on the sample surface at a glancing angle of 15 degrees from horizontal. The lamp spectrum was filtered using a 280 nm low pass filter, corresponding to a photon energy of 4.4 eV. The final image was projected onto a microchannelplate image intensifier and recorded with a video camera. Pictures captured in Ohio University also show that indented areas appear brighter, corresponding to more electron emission.

## Conclusions

New quantum interference effect in solids has been predicted theoretically and observed experimentally by several groups (TSU, PSU, Nanostructures Inc. Ohio University). Work function decrease was observed in such materials as Au and $SiO_2$.

Experimental results are in good qualitative agreement with predictions of the theory. In order to achieve quantitative agreement, structure of the solid should be made single crystal or close to ideally amorphous that has not been realize in experiments so far, and is the subject of future experiments. It is expected that magnitude of the effect will increase dramatically, with improving structure of the solid and reducing the width of indents. First could be realized by improving thin film deposition technology and reducing the contamination. Second could be achieved by moving to more advanced technologies such as e-beam and ion- beam lithography. Magnitude of the effect should be higher for solids with low value of Fermi energy and low value of work function. For example such a material as $LaB_6$ is expected to allow work functions less than 0.5 eV. In that case it will find application in thermionic and thermotunnel refrigerators and energy generators operating at room temperature.

## Acknowledgments

We thank Rezo Guliaevi for X-ray diffraction analysis and Stuart Harbron for reading the text. The work is financed and supported by Borealis Technical Limited, assignee of corresponding patents (US 6,281,514; US 6,495,843; US 6,680,214; US 6,531,703; US 6,117,344) and all provisional and pending patent applications. All intellectual property has been licensed to Avto Metals plc, Cool Chips plc, or Power Chips plc.